\documentclass[12pt]{article}

\begin{document}
\title{Renormalization Invariants of the Neutrino Mass Matrix}
\author{Sanghyeon Chang\footnote{schang@physics.purdue.edu} \ and
T.~K.~Kuo\footnote{tkkuo@physics.purdue.edu} \\ \small\it
Physics. Dept. Purdue University, West Lafayette, IN 47906, U.S.A.}
\maketitle
\abstract{The renormalization evolution of all parameters in the neutrino mass
matrix depends only on one variable, the energy scale. This fact, coupled with
rephasing considerations, leads to a set of renormalization invariants,
correlating the evolution of physical parameters. For the general three flavor
case, we obtain these invariants
explicitly and discuss their implications.
\section{Introduction}
\baselineskip 6mm
Recent results from the  atmospheric and solar experiments have shown strong 
evidence for neutrino oscillations.~\cite{superK,cleveland}
These observations indicate
that neutrinos are not massless, and that two of the
neutrino mixing angles are large, or even maximal. This is in contrast to the quark
mixing angles, which are all small. Even with our limited knowledge, it seems
clear that the neutrino mass matrix, just like their quark counterpart, has a
rich structure, and it is urgent to have an understanding of its salient
features.  To account for the minuscule neutrino masses, the seesaw model
\cite{gellmann} makes use of a
heavy scale for the right-handed neutrinos. Thus, any theoretical
understanding of the observed neutrino parameters necessarily involves two
vastly different energy scales. This means that renormalization effects must
be taken into account in any theoretical model of the neutrino mass matrix.

The renormalization of the neutrino mass matrix has been extensively discussed
in the literature \cite{antusch,casas,kuo1}. Although different models, such as the SM or MSSM, give rise to numerically distinct
results for individual parameters, as we have shown for the two flavor
problem \cite{kuo1},
there are renormalization group equation (RGE) invariants, correlating the
evolution of the physical parameters. These invariants are the consequences of
the general structure of the RGE, and remain the same for any model which
conserves FCNC.

In general, RGE evolution implies that each physical parameter becomes a
function of the energy scale.
Thus, if there are $n$ independent parameters in the mass matrix, we might expect to
have $(n-1)$ RGE invariants. However, the mass
matrices are also subject to arbitrary rephasing transformations. In addition,
such phases are generated by the renormalization transformation. As a result,
the physical RGE invariants must also be rephasing invariants. For the three
flavor mass matrix, it turns out that there are three (complex) RGE and
rephasing invariants, amongst its eight physical parameters. These invariants
will be detailed in Sec.~4. Just as for the two flavor problem, these
invariants are universal, valid for a wide class of models.

\section{Parameterization of Mass Matrices}

Before we embark on a detailed discussion of renormalization, it is useful to
introduce a general parameterization of the mass matrix which will facilitate
the analysis.
In this paper, we will consider the (symmetric) neutrino mass matrix to
originate from a dimension five term in the effective Lagrangian,
\begin{equation}
L = f\nu^T\nu\langle\phi\rangle\langle\phi\rangle
=\nu^T M^0_\nu \nu .
\end{equation}
Here, $\langle\phi\rangle$ is the vacuum expectation value of the Higgs scalar, $f$ is the
coupling constant, $M^0_\nu$ is the neutrino mass matrix, and $\nu$ is the
neutrino wave function in the flavor basis.

We can write, in general,
\begin{equation}
M^0_\nu = U M^{\rm diag} U^T, \label{mass2}
\end{equation}
\begin{eqnarray}
M^{\rm diag} &=&\left( \begin{array}{ccc} e^{2 \eta_1} & & \\ & e^{2 \eta_2} & \\
 & & e^{2\eta_3} \end{array} \right) , \label{mass} \\
U &=& P e^{-i\epsilon_7\lambda_7}e^{-i\epsilon_5\lambda_5}e^{-i\epsilon_3\lambda_3}
e^{-i\epsilon_2\lambda_2} P' ,\\
P &=&\left( \begin{array}{ccc}e^{i \alpha_1} & & \\ & e^{i \alpha_2} & \\
 & & e^{i \alpha_3} \end{array} \right) , \\
P' &=&\left( \begin{array}{ccc} e^{i \gamma_1} & & \\ & e^{i \gamma_2} & \\
 & & e^{i \gamma_3} \end{array} \right) . 
\end{eqnarray}
Here, the mass eigenvalues are given by $\exp(2\eta_i)$, $\alpha_i$ are the
unphysical phases from the neutrino wave functions, $(\epsilon_2,
\epsilon_5,\epsilon_7)$ are the
physical neutrino mixing angles, $\epsilon_3$ is a CP violating phase, the
$\gamma$'s are the intrinsic CP phases of the mass eigenvalues, and the
$\lambda$'s are the Gell-Mann matrices. To preserve the symmetry of the flavors,
we will not use the diagonal $\lambda$'s in $M^{\rm diag}$.
Note also that, since neutrino oscillations are governed by the effective
Hamiltonian, $H = (M_\nu M_\nu^\dagger)/2E$, they are independent of the
phases $\gamma_i$.

It is convenient to factor out the determinant (the overall scale) of $M^0_\nu$
and define
\begin{eqnarray}
M_\nu &=& (det M^0_\nu)^{-1/3} M^0_\nu ,\\
det M_\nu&=&1 .
\end{eqnarray}
The condition
$det M_\nu =1$ is obtained by imposing the following relations on $M^0_\nu$:
\begin{eqnarray}
\sum \eta_i =\sum \alpha_i =\sum \gamma_i =0 .
\end{eqnarray}
This corresponds to the fact that, with $det M_\nu =1$, $M_\nu$ only depends
on the mass ratios $(\Delta \eta)$ and the relative phases $(\Delta \alpha$
and $ \Delta \gamma)$. Note that, with the parametrization in Eq.(\ref{mass}),
$M_\nu$ can be analytically continued into an $SU(3)$ matrix
$(\eta_j\rightarrow i\eta_j)$. A fact which will be used later. Also,
rephasing of the neutrino wave functions changes only $\alpha_j$, while
leaving all other parameters invariant.

It is also useful to write down the symmetric matrix $M_\nu$ explicitly.
\begin{eqnarray}  
M_\nu =\left(\begin{array}{lll}
c_{(5)}^2 \bar{\chi}_1 + s_{(5)}^2 e^{2\bar{\eta}_3} &
\begin{array}{l}
c_{(7)} c_{(5)} s_{2(2)} \Delta_{12} \\ - s_{(7)}s_{2(5)}\Delta_{13} 
\end{array}& 
\begin{array}{l}
s_{(7)} c_{(5)} s_{2(2)} \Delta_{12} \\+ c_{(7)}s_{2(5)}\Delta_{13}  
\end{array}\\[5mm]
\begin{array}{l}
c_{(7)} c_{(5)} s_{2(2)} \Delta_{12} \\ - s_{(7)}s_{2(5)}\Delta_{13}
\end{array}
&\begin{array}{l} c_{(7)}^2 \bar{\chi}_2 -  s_{2(7)} s_{(5)}s_{2(2)}\Delta_{12} \\ + s_{(7)}^2 ( s_{(5)}^2 \bar{\chi}_1 +
c_{(5)}^2 e^{2\bar{\eta}_3})  
\end{array}& 
\begin{array}{l}  
s_{2(7)}(  \bar{\chi}_2 - s_{(5)}^2 \bar{\chi}_1 +
c_{(5)}^2 e^{2\bar{\eta}_3})/2 \\ +  c_{2(7)} s_{(5)}s_{2(2)}\Delta_{12}
\end{array}
\\[5mm]
\begin{array}{l}
s_{(7)} c_{(5)} s_{2(2)} \Delta_{12} \\+ c_{(7)}s_{2(5)}\Delta_{13}
\end{array}
& \begin{array}{l}
s_{2(7)}(\bar{\chi}_2-s_{(5)}^2 \bar{\chi}_1 +
c_{(5)}^2 e^{2\bar{\eta}_3})/2 \\ +  c_{2(7)} s_{(5)}s_{2(2)}\Delta_{12}
\end{array}
& \begin{array}{l} 
s_{(7)}^2 \bar{\chi}_2+  s_{2(7)} s_{(5)}s_{2(2)}\Delta_{12} \\ + c_{(7)}^2 ( s_{(5)}^2 \bar{\chi}_1 +
c_{(5)}^2 e^{2\bar{\eta}_3})  
\end{array}
\end{array}
\right) ,  \label{matrix1}
\end{eqnarray}  
\begin{eqnarray}  
\bar{\eta}_i& =& \eta_i + i \gamma_i,\\
\chi_1&=& c_{(2)}^2 e^{2\bar{\eta}_1} + s_{(2)}^2 e^{2\bar{\eta}_2},
\ \ \ \bar{\chi}_1= e^{-2i\epsilon_3}\chi_1,\\
\chi_2&=& s_{(2)}^2 e^{2\bar{\eta}_1} + c_{(2)}^2 e^{2\bar{\eta}_2},
\ \ \ \bar{\chi}_2= e^{2i\epsilon_3}\chi_2,\\
\Delta_{12} &=&\frac{1}{2} \left( e^{2\bar{\eta}_1} -  e^{2\bar{\eta}_2}\right),
\label{condition2}\\
\Delta_{13} &=&\frac{1}{2} \left( \bar{\chi}_1 -  e^{2\bar{\eta}_3}\right).
\label{condition1}
\end{eqnarray}  
Here, we use the notation $s_{(2)}= \sin \epsilon_2, s_{2(2)}=\sin 2\epsilon_2$,
etc. We have also set $\alpha_i=0$, without loss of generality.

\section{RGE}

The RGE for the effective neutrino mass matrix, $M_\nu$, has been studied
extensively. In the SM and MSSM, the RGE were obtained explicitly and can be
written in
the form \cite{antusch, casas, kuo1}
\begin{equation}
\frac{d}{dt} M^0_\nu = \kappa M^0_\nu +\left\{ Q, M^0_\nu\right\},
\end{equation}
where $\kappa$ is a constant, $Q$ is a diagonal and traceless matrix 
\begin{eqnarray}
tr Q =0 .
\end{eqnarray}
 $t$ is the scale variable $t=\frac{1}{16\pi^2}\ln (\mu/\mu_0)$,
with $(\mu,\mu_0) =$ energy scale. The solution is given by
\begin{eqnarray}
det M^0_\nu(t)&=& e^{3\kappa t} det M^0_\nu (0), \\
M_\nu(t) &=& e^{Qt} M_\nu(0) e^{Qt} . \label{reno}
\end{eqnarray}
 The quantities $\kappa$ and $Q$ were given explicitly in terms of the leptonic
Yukawa constants.

We note that the form of Eq.(\ref{reno}) is quite general. The operator 
$e^{Qt}$ originates from the (relative) neutrino wave function
renormalization. It amounts to a change of relative scale (rescaling) between
the different flavors. As long as the interactions responsible for the
renormalization do not contain FCNC, their effect will be a diagonal matrix
multiplication, as in Eq.(\ref{reno}). The close relation between rescaling
(renormalization) and rephasing is revealed by considering pure imaginary
values for $Q$, which turns Eq.(\ref{reno}) into a rephasing transformation.
If, in addition, we consider $\eta_j \rightarrow i\eta_j$ in $M_\nu$, then the
equation becomes a rephasing transformation in $SU(3)$.

Eq.(\ref{reno}) is a formal solution of RGE, since it only gives the
$t$-dependence of the matrix elements of $M_\nu$. One would really like to know
the $t$-dependence of the physical parameters. To this end, we must reexpress
$M_\nu$ in Eq.(\ref{reno}) in the form of Eq.(\ref{mass2}),
\begin{eqnarray}
M_\nu (t) = U(t) M^{\rm diag} (t) U^T(t), \label{reno2}
\end{eqnarray}
and one need to relate the physical parameters at scale $t$ to those at $t=0$.

Mathematically, the RGE solution corresponds to the 
relation connecting the parameters, known as
the Baker-Campbell-Hausdorff (BCH) formula, between different rearrangements
(e.g. Eqs.(\ref{reno}) and (\ref{reno2}))
of the non-commuting factors in an element of  $SL(3,C)$.
Since we are dealing with only exponential functions which are free of
singularities, these relations should remain valid under analytic
continuations. In particular, the BCH formulae for $SL(3,C)$ are the analytic
continuation of those for $SU(3)$. These ideas can be implemented explicitly
for the case of two flavors, which we will study first before we take on the
full analysis of the three flavor problem.

Consider a general symmetric $SU(2)$ matrix (with two real parameters), which
can be written
as
\begin{eqnarray}
 \widetilde{N_1} =  e^{-i\beta\sigma_2}  e^{2i\tilde{\gamma}\sigma_3} e^{i\beta\sigma_2},
\label{para1}
\end{eqnarray}
or 
\begin{eqnarray}
 \widetilde{N_2} = e^{i\tilde{\omega}\sigma_3}  e^{2i\tilde{\tau}\sigma_1}e^{i\tilde{\omega}\sigma_3}.
\label{para2}
\end{eqnarray}
The relations (BCH formula) between the two parametrizations $\widetilde{N_1}$
and $\widetilde{N_2}$
can be read off from the matrix elements as given in Eqs (\ref{para1}), (\ref{para2}).
\begin{eqnarray}
\cos{2\tilde{\gamma}} &=  &\cos{2\tilde{\tau}}\cos{2\tilde{\omega}},  \\
\sin{2\tilde{\gamma}}\cos{2\beta} &=  &\cos{2\tilde{\tau}}\sin{2\tilde{\omega}},  \\
\sin{2\tilde{\gamma}}\sin{2\beta} &=  &\sin{2\tilde{\tau}} . 
\end{eqnarray}
When $\widetilde{N}$ is subject to a rephasing transformation, 
\begin{equation}
\widetilde{N}\rightarrow
e^{-i\tilde{\alpha}\sigma_3}  \widetilde{N} e^{-i\tilde
{\alpha}\sigma_3},  \label{rephas}
\end{equation}
it is obvious that parametrization Eq.(\ref{para2}) makes it
trivial, resulting in $\tilde{\omega} \rightarrow \tilde{\omega}- \tilde{\alpha}$.
But for  $\widetilde{N_1}$ in Eq.(\ref{para1}), it induces the change $\beta \rightarrow \beta', \tilde{\gamma}\rightarrow
\tilde{\gamma}'$, satisfying
\begin{eqnarray}
\cos{2\tilde{\gamma}'} &=
&\cos{2\tilde{\tau}}\cos{2(\tilde{\omega}-\tilde{\alpha})},  \label{cos}\\
\sin{2\tilde{\gamma}'}\cos{2\beta'} &=  &\cos{2\tilde{\tau}}\sin{2(\tilde{\omega}-\tilde{\alpha})},  \\
\sin{2\tilde{\gamma}'}\sin{2\beta'} &=  &\sin{2\tilde{\tau}} .
\end{eqnarray}
It follows that 
\begin{equation}
\tan 2\beta' = \frac{ \sin{2\beta}/ \cos{2\tilde{\alpha}}}
{\cos{2\beta} - \tan 2\tilde{\alpha}/ \tan 2\tilde{\gamma}} . \label{tan}
\end{equation}
In addition, $\tilde\tau$ is invariant under rephasing. Thus, we have the rephasing
invariant, in terms of the parametrization in Eq.(\ref{para1}).
\begin{equation}
\sin{2\tilde{\gamma}}\sin{2\beta} = \sin{2\tilde{\gamma}'}\sin{2\beta'}
\label{sin} .
\end{equation}
Eqs.~(\ref{tan}) and (\ref{sin}) are the solutions to the $SU(2)$
rephasing transformation, Eq.(\ref{rephas}).

The same results can be taken over for symmetric mass matrices, where all
variables $(\beta, \tilde{\gamma},\tilde{\omega},\tilde{\tau})$ are complex,
with four real parameters. Let us first consider the case of real mass
matrices,  corresponding to pure
imaginary $(\tilde{\gamma},\tilde{\omega},\tilde{\tau})$ :
\begin{equation}
(\tilde{\gamma},\tilde{\omega},\tilde{\tau}) \longrightarrow (i \gamma, i
\omega, i \tau).
\end{equation}
The resulting mass matrix can be written in two alternative forms,
\begin{eqnarray}
N_1 & = & e^{-i\beta\sigma_2}  e^{-2\gamma\sigma_3} e^{i\beta\sigma_2}, \nonumber\\
N_2 & = & e^{-\omega\sigma_3}  e^{-2\tau\sigma_1}e^{-\omega\sigma_3} .
\end{eqnarray}
With $\tilde{\alpha} \rightarrow i\alpha$, the rephasing transformation on
$\widetilde{N}$ becomes a renormalization (rescaling) transformation on $N$ :
\begin{equation}
 N \longrightarrow e^{\alpha \sigma_3} N e^{\alpha \sigma_3}. \label{rescale}
\end{equation}
The solution to the RGE is obtained directly from the BCH formula of $SU(2)$,
Eq.(\ref{tan}), by analytic continuation:
\begin{equation}
\tan 2\beta' = \frac{ \sin{2\beta}/ \cosh{2\alpha}}
{\cos{2\beta} - \tanh 2{\alpha}/ \tanh 2\gamma} . \label{tan2}
\end{equation}
This is the same relation obtained earlier for RGE evolution. At the same
time, we have an RGE invariant \cite{kuo1} :
\begin{equation}
 \sin 2\beta \sinh 2\gamma = \sin 2\beta' \sinh 2\gamma'. \label{inv}
\end{equation}
Note that this is also automatically invariant under (relative) rephasing on
the mass matrix $N$.
In addition, while the value $\alpha$ in Eq.(\ref{rescale}) is model dependent,
Eq.(\ref{inv}) is not. 

In general, all parameters $(\beta, \tilde{\gamma},\tilde{\omega},\tilde{\tau})$
are complex, so that $\widetilde{N_1}$ and $\widetilde{N_2}$ become complex
mass matrices. As was shown in Ref.\cite{kuo2}, we can demand that $\beta$ and
$\beta'$ be real by adding rephasing factors to $\widetilde{N_1}$. 
 The resulting generalization of Eq.(\ref{tan}) coincides
with Eq.(15) of Ref.\cite{kuo1}. In this connection, we emphasize that the renormalization
transformation, Eq.(\ref{rescale}), generates rephasing factors when
$N$ is complex, as was shown explicitly in Ref.\cite{kuo1}.

Schematically, solutions to the rescaling transformations on mass matrices are
the BCH formulas in $SL(2,C)$, which can be obtained from the rephasing
transformations in $SU(2)$, by analytic continuation. We can represent this in
a commutative diagram: 
\begin{equation}
\begin{array}{ccc}
\widetilde{N}_1 & \stackrel{\mathrm{BCH }\ SU(2)}{\longleftarrow\hspace{-2mm} -\hspace{-2mm}\longrightarrow} & \widetilde{N}_2 \\
\vcenter{\llap{$ \tiny\begin{array}{l}\mathrm{Analytic}\\
\mathrm{continuation}\end{array} $}} 
\Big\updownarrow & & \Big\updownarrow 
\vcenter{\rlap{$ \tiny\begin{array}{l}\mathrm{Analytic}\\
\mathrm{continuation}\end{array} $}}  \\
N_1 & \stackrel{\mathrm{BCH}\ SL(2,C)}
{\longleftarrow\hspace{-2mm} -\hspace{-2mm}\longrightarrow}
& N_2
\end{array}
\end{equation}

We now turn to the case of three flavors.
A general symmetric $SU(3)$ matrix (with five parameters) can be parametrized in
either of two ways
\begin{eqnarray}
 W_1 &=  &V\, W^{\rm diag}\, V^T,  \\
 W^{\rm diag} &=  & \left( \begin{array}{ccc} e^{i2 \eta_1} & & \\ & e^{i2
\eta_2} & \\ & & e^{i2\eta_3} \end{array} \right) , \mbox{  } \sum\eta_i =0 \\
V &=& 
e^{-i\epsilon_7\lambda_7}e^{-i\epsilon_5\lambda_5} e^{-i\epsilon_2\lambda_2}  ,
\end{eqnarray}
or
\begin{eqnarray}
 W_2 &=  & P e^{i(\xi_1\lambda_1 +\xi_4\lambda_4 + \xi_6 \lambda_6)}P,  \\
P &=&\left( \begin{array}{ccc}e^{i \delta_1} & & \\ & e^{i \delta_2} & \\
 & & e^{i \delta_3} \end{array} \right) ,  \mbox{  } \sum\delta_i =0 .
 \label{P}
\end{eqnarray}
The order of the non-commuting matrix products are chosen so that $W_1$
corresponds to the usual mass matrix parametrization, while $W_2$ is most
convenient for rephasing considerations.

The BCH formulas yield analytic relations between the two sets of parameters
$(\eta_i, \epsilon_j)$ and $(\xi_i, \delta_j)$. A rephasing transformation
would change $\delta_j \rightarrow \delta'_j$, and the corresponding
transformation on $(\eta_i, \epsilon_j)$ can be calculated as in
Eqs.(\ref{cos}-\ref{tan}).
Also, the functions $\xi_i$ in terms of $(\eta_i, \epsilon_j)$ are rephasing
invariants. The functions for $(\eta_i, \epsilon_j)$ obtained would have
provided explicit solutions to the RGE, as in Eq.(\ref{tan2}). Unfortunately,
owing to the complexity of the $SU(3)$ algebra, so far we are unable to solve
for these functions explicitly.

When we let all parameters assume complex values, the matrix $W$ turns into a
symmetric mass matrix. A rescaling (renormalization) transformation corresponds
to $\delta_j \rightarrow \delta_j + i\Delta_j$.
Thus, the rescaling (RGE) invariants which are also rephasing invariants are
precisely $(\xi_1, \xi_4, \xi_6)$, as complex functions of $(\eta_i,
\epsilon_j)$.
Although we can not obtain these functions explicitly, we can obtain the RGE
invariants using the matrix elements as the variables. As before, we can
arrive at the results by first studying the rephasing invariants in $SU(3)$.

Consider rephasing transformations
on a general $SU(3)$ matrix (with eight parameters), which can be written in the form
\begin{equation}
 V = P
e^{-i\epsilon_7\lambda_7}e^{-i\epsilon_5\lambda_5}e^{-i\epsilon_3\lambda_3}
e^{-i\epsilon_2\lambda_2} P' ,
\end{equation} 
where $P$ is given in Eq.(\ref{P}) and $P'$ is obtained from $P$ by the
substitution $\delta_j \rightarrow \delta_j'$.
This is precisely the parametrization for the CKM matrix and
$(\epsilon_2,\epsilon_5,\epsilon_3, \epsilon_7)$ are rephasing invariants. In
terms of the matrix elements of $V$, $V_{IJ}$, rephasing transformation gives
$V_{IJ}\rightarrow e^{i(\delta_I+\delta'_J)}V_{IJ}$. Thus, the rephasing
invariants are $|V_{IJ}|^2$. When we let the parameters be complex, $V$
analytically continues into $M$, the mass matrix. The rephasing and rescaling
invariants are now given by $M^{-1}_{IJ} M_{IJ}$.

Besides $|V_{IJ}|^2$, another familiar form of the rephasing invariant
\cite{jarlskog} is
given by  ${\cal
J}_{IJKL}= V_{IJ}V_{KL}V^*_{IL}V^*_{KJ}$. When we impose the condition $det V
=1 $, only relative rephasings, but no overall phases, are admitted. The
rephasing invariant takes a simpler form
\begin{equation}
 I_s = e_{IJK} e_{I'J'K'} V_{II'}V_{JJ'}V_{KK'}.
\end{equation}
There are six different ways to arrange the indices so that we may label $I$
by $s$, which is an element of the permutation group $S_3$,
$$
s=\left(\begin{array}{ccc} I & J & K \\ I' & J' & K'\end{array} \right),
\mbox{ denoting } ( I \rightarrow I', J \rightarrow J', K  \rightarrow K').
$$
Note that
\begin{equation}
\sum_s I_s = det V =1 .
\end{equation}
Also, $I_s$ has a simple relation to the familiar rephasing invariant ${\cal
J}_{IJKL}$. For a unitary $V$ with $det=1$, its
minors are just the complex conjugated elements. For instance, 
\begin{equation}
 V_{11} = V^*_{22}V^*_{33} - V^*_{23}V^*_{32},
\mbox{ or }
V_{11}V_{22}V_{33} = |V_{22}|^2|V_{33}|^2- V_{22}V_{33}V^*_{23}V^*_{32}.
\end{equation}
When we let the indices take on different values, it is easy to show that all
of the products $I_s$ are similarly related to ${\cal J}_{IJKL}$, with Im$I_S= -
$Im$({\cal J}_{IJKL})$, independent of the indices.

The analytic continuation of the rephasing invariants $I_s$ turns them into
rescaling and rephasing invariants  for the mass matrices.
\begin{equation}
 {\cal J}_s = e_{IJK}e_{LMN}M_{IL}M_{JM}M_{KN} .
\end{equation}
\section{RGE Invariants}
As we discussed in the previous section, the physical RGE invariants are the rephasing
and rescaling invariants of the mass matrix. There are two equivalent forms of
these invariants. 1) $M^{-1}_{IJ}M_{IJ}$, 2)
$e_{IJK}e_{LMN}M_{IL}M_{JM}M_{KN}$ . When we convert these into physical
variables, it turns out that the former is more convenient, which will be
presented in the following.

Let us define
\begin{equation}
I_{IJ} =M^{-1}_{IJ}M_{IJ}.
\end{equation}
These invariants are not independent, since $M_{IJ}=M_{JI}$, and
\begin{equation}
 \sum_I I_{IJ}= \sum_J I_{IJ} =1 .
\end{equation}
 So there are altogether three independent (complex) invariants, which we can
take to be
\begin{equation}
 I_1=I_{11}-1, I_2=I_{12}-I_{13} \mbox{ and } I_3=I_{23}. 
\end{equation}
Explicitly, we find
\begin{eqnarray}
 I_1 &=  &M_{11}M^{-1}_{11} -1  \nonumber\\
 &=  &  c^4_{(5)} s^2_{2(2)} \sinh^2(\overline\eta_1-\overline\eta_2) 
+ s^2_{2(5)} c^2_{(2)} \sinh^2(\overline\eta_1-\overline{\overline{\eta}}_3 )\nonumber\\  &&
+ s^2_{2(5)} s^2_{(2)} \sinh^2(\overline\eta_2-\overline{\overline{\eta}}_3) ,\nonumber\\
I_2  &=  &M^{-1}_{12}M_{12} - M^{-1}_{13}M_{13}\nonumber\\
&=& s_{2(2)}s_{2(5)}c_{(5)}s_{2(7)} \left(\Delta_{13}\Delta^-_{12}
+\Delta^-_{13} \Delta_{12}\right) \nonumber\\ && + c_{2(7)}\left(s^2_{2(2)}c^2_{(5)}
\Delta_{12}\Delta^-_{12} -s^2_{2(5)}
\Delta_{13}\Delta^-_{13}\right),\label{invariant}\\
I_3  &=  &M_{23}M^{-1}_{23} \nonumber\\
&=& \left[ s_{2(7)}\left(\bar\chi_2 - s^2_{(5)}\bar\chi_1 -
c^2_{(5)}e^{2\overline\eta_3}\right)/2 + s_{2(2)}s_{(5)}c_{2(7)} \Delta_{12}\right]
\nonumber\\
&& \times \left[ s_{2(7)}\left(\bar\chi^-_2 - s^2_{(5)}\bar\chi_1^- -
c^2_{(5)}e^{-2\overline\eta_3}\right)/2 + s_{2(2)}s_{(5)}c_{2(7)}
\Delta^-_{12}\right] .
\nonumber
\end{eqnarray}
Here we have used the notations in Eqs.(\ref{matrix1}-\ref{condition1}). In addition,
\begin{eqnarray}
 \overline{\overline{\eta}}_3 &=  & e^{2i\epsilon_3}\overline{\eta}_3 , \\
\Delta^-_{12}  & = & \frac{1}{2}\left(e^{-2\overline{\eta}_1}- e^{-2\overline{\eta}_2}
\right),\\
\Delta^-_{13}  & = & \frac{1}{2}\left(\bar{\chi}^-_1- e^{-2\overline{\eta}_3}
\right),\\
\bar{\chi}^-_1 &=&  e^{2i\epsilon_3} \left(c^2_{(2)}e^{-2\overline{\eta}_1}
+ s^2_{(2)}e^{-2\overline{\eta}_2}\right),\\
\bar{\chi}^-_2 &=&  e^{-2i\epsilon_3} \left(s^2_{(2)}e^{-2\overline{\eta}_1}
+ c^2_{(2)}e^{-2\overline{\eta}_2}\right).
\end{eqnarray}

These invariants show that the physical parameters are intricately
correlated during the RGE evolution. In general, we can not single out one
or two variables which evolve independently of the others. However, in limited
regions when certain conditions are satisfied, we do get simplified relations
between a subset of the parameters. We will highlight some of these relations
in the following.
\\
1)  Real Mass matrix.  

In this case, all physical phases vanish, $\epsilon_3=\gamma_i =0$. The RGE
invariants $I_{1,2,3}$ are all real so that renormalization does not generate
any physical phases, as expected from Eq.(\ref{reno}).
\\
2) Two flavor solutions.

The three flavor problem reduces to that of two flavor under certain
conditions. This happens when two of the three mixing angles vanish. Thus, if
$s_{(5)}=s_{(7)}=0$, we find that $I_1=I_2=s_{2(2)}^2 \sinh^2(\overline{\eta}_1
-\overline{\eta}_2), I_3=0$. Note that the condition $s_{(5)}=s_{(7)}=0$ is RGE
stable. Similarly, if $s_{(2)}=s_{(7)}=0$, or if $s_{(2)}=s_{(5)}=0$, the
result are genuine two flavor solutions.
\\
3) In regions when one angle is small.

It may happen that, in a certain range of $t$, one of the mixing angles can be
small. In general, such conditions can only be fulfilled in a limited region.
There will then be approximate invariant combinations from a reduced set of
parameters. For instance, in a region where $s_{(2)}\rightarrow 0$, we find
($\overline{\overline{\eta}}_1 =\overline{\eta}_1 - i\epsilon_3,
\overline{\overline{\eta}}_2 =\overline{\eta}_2 + i\epsilon_3 $),
\begin{eqnarray}
I_1&\rightarrow  &s_{2(5)}^2 \sinh^2(\overline{\eta}_1 -\overline{\eta}_3)\nonumber\\
I_2&\rightarrow  & -c_{2(7)}s_{2(5)}^2 \sinh^2(\overline{\overline{\eta}}_1
-\overline{\eta}_3) \\
I_3&\rightarrow  & s_{2(7)}^2\left[-s_{(5)}^2
\sinh^2(\overline{\overline{\eta}}_2 -\overline{\overline{\eta}}_1)
-c_{(5)}^2 \sinh^2(\overline{\overline{\eta}}_2 -\overline{\eta}_3)
+s_{(5)}^2c_{(5)}^2 \sinh^2(\overline{\overline{\eta}}_1 -\overline{\eta}_3)
\right]. \nonumber 
\end{eqnarray}
This means that, in the limit $s_{(2)}\rightarrow 0$, the $(1-3)$ sector
behaves like a two flavor problem. At the same time, there is a correlation
between the mixing angle $\epsilon_7$ and the CP phase, $\epsilon_3$. However,
if $\epsilon_3=0$, then the approximation $s_{(2)}\rightarrow 0$ is consistent
only if $\epsilon_7\rightarrow 0$. Similar conclusions can be reached for the
case $s_{(5)}\rightarrow 0$ and $s_{(7)}\rightarrow 0$. In addition, the cases
when two masses are nearly degenerate, or when one mass value dominates, can
also be analyzed along these lines.
\section{Conclusion}
In this paper, we have studied the properties of the three flavor neutrino
mass matrix under RGE evolution. Unlike the two flavor problem, where we
obtained exact analytic solutions, the algebra of the $3\times 3$ matrices is
formidable, and we are only able to find three (complex) RGE invariants which
correlate the evolutions of the many physical parameters.

The RGE evolution of a (symmetric) mass matrix with $det M=1$, for FCNC
conserving theories (including the SM and MSSM), is given in the form
$M\rightarrow e^Q Me ^Q$, $Q=$ real diagonal and $tr Q =0$. That is,
renormalization amounts to a relative rescaling between the different flavors.
At the same time, $M$ is subject to arbitrary rephasing transformations, which
correspond to taking $Q$ to be an arbitrary pure imaginary matrix. 
The combined rescaling and rephasing transformation is thus given by
$M\rightarrow e^{\bar{Q}} Me^{\bar{Q}}$,  $\bar{Q}=$ complex. Since $\bar{Q}$ contains
only two (complex) variables, we expect RGE invariants formed from the many
physical variables in $M$. By means of an analytic continuation, these
considerations are the same as those in obtaining rephasing invariants of the
CKM matrix. We can thus write down three (complex) RGE and rephasing
invariants explicitly. Our arguments also make it clear that these invariants
are universal, independent of any specific model used to calculate the RGE
evolution. 

Since exact solutions for the three flavor problem are not available, a number
of approximate solutions have been considered in the literature \cite{casas}. The RGE invariants can be used
to check the consistency of these approximations and to suggest viable new
ones. The structure of these invariants also shows that, while the two flavor
approximation is natural in a number of situations, their validity can only be
established for a limited range of $t$. For large $t$, when the parameters
also vary considerably, the two flavor approximation is viable only under
very stringent conditions.

With minor changes, most of the arguments in this work can be adapted to the
study of quark mass matrices. In fact, \cite{kuo1} it was shown that the infrared
fixed point for two flavor RGE evolution corresponds to $\beta \rightarrow 0,
m_2/m_1 \rightarrow \infty$. The approach to the fixed point, however, is
governed by  Eq.(\ref{inv}), giving  $\beta \sqrt{m_2/m_1}\rightarrow$
constant. This suggests that the quark mass matrices are the results of
large RGE evolution, and that the well-known empirical relations between
mixing angle and mass rations, $\theta_{ij}\sim\sqrt{m_i/m_j}$, may have a
dynamical origin. We plan to apply our analysis to a detailed study of the quark sector in the
future.

This work is supported in part by DOE grant No.
DE-FG02-91ER40681.

\end{document}